\documentclass[aps,pra,reprint,superscriptaddress,floatfix]{revtex4-1}
\usepackage[T1]{fontenc}
\usepackage[utf8]{inputenc}
\usepackage{multirow}
\usepackage{braket}
\usepackage{bbm}
\usepackage{bm}
\usepackage{booktabs}
\usepackage{amsmath}
\usepackage{amsfonts}
\usepackage{amsthm}
\usepackage{mathtools}
\DeclarePairedDelimiter{\ceil}{\lceil}{\rceil}

\theoremstyle{definition}

\usepackage{graphicx}
\usepackage{placeins}
\usepackage{xcolor}
\usepackage[]{hyperref}
\hypersetup{
   pdftitle={Entanglement bounds on the performance of quantum computing architectures},
    colorlinks=true,
    citecolor=blue
}
\usepackage[all]{hypcap}
\usepackage[normalem]{ulem}

\newcommand{\abs}[1]{\ensuremath{\left| #1 \right|}}
\newcommand{\half}{\frac{1}{2}}

\newcommand{\rb}{\mathrm{RB}}

\newcommand{\whp}[0]{\mathbin{\Pi_\alpha}}
\newcommand{\vhp}[0]{\mathbin{\Pi_{\vec{\alpha}}}}
\newcommand{\uhp}[0]{\mathbin{\Pi}}

\newcommand{\C}{\mathcal{C}}

\usepackage[capitalise,compress]{cleveref}
\crefname{section}{Sec.}{Secs.}
\Crefname{section}{Section}{Sections}
\crefrangelabelformat{equation}{\textup{(#3#1#4)}--\textup{(#5#2#6)}}

\begin{document}
\title{Entanglement bounds on the performance of quantum computing architectures}
\date{June 30, 2020}

\author{Zachary Eldredge}
\affiliation{Joint Center for Quantum Information and Computer Science, NIST/University of Maryland, College Park, Maryland 20742, USA}
\affiliation{Joint Quantum Institute, NIST/University of Maryland, College Park, Maryland 20742, USA}
\author{Leo Zhou}
\affiliation{Department of Physics, Harvard University, Cambridge, Massachusetts 02138, USA}
\author{Aniruddha Bapat}
\affiliation{Joint Center for Quantum Information and Computer Science, NIST/University of Maryland, College Park, Maryland 20742, USA}
\affiliation{Joint Quantum Institute, NIST/University of Maryland, College Park, Maryland 20742, USA}
\author{James R. Garrison}
\affiliation{Joint Center for Quantum Information and Computer Science, NIST/University of Maryland, College Park, Maryland 20742, USA}
\affiliation{Joint Quantum Institute, NIST/University of Maryland, College Park, Maryland 20742, USA}
\author{Abhinav Deshpande}
\affiliation{Joint Center for Quantum Information and Computer Science, NIST/University of Maryland, College Park, Maryland 20742, USA}
\affiliation{Joint Quantum Institute, NIST/University of Maryland, College Park, Maryland 20742, USA}
\author{Frederic T. Chong}
\affiliation{Department of Computer Science, University of Chicago, Chicago, Illinois 60637, USA}
\author{Alexey V. Gorshkov}
\affiliation{Joint Center for Quantum Information and Computer Science, NIST/University of Maryland, College Park, Maryland 20742, USA}
\affiliation{Joint Quantum Institute, NIST/University of Maryland, College Park, Maryland 20742, USA}

\begin{abstract}
There are many possible architectures of qubit connectivity that designers of future quantum computers will need to choose between. However, the process of evaluating a particular connectivity graph's performance as a quantum architecture can be difficult.
In this paper, we show that a quantity known as the isoperimetric number establishes a lower bound on the time required to create highly entangled states.
This metric we propose counts resources based on the use of two-qubit unitary operations, while allowing for arbitrarily fast measurements and classical feedback.
We use this metric to evaluate the hierarchical architecture proposed by A.\ Bapat \textit{et al.} [Phys.\ Rev.\ A \textbf{98}, 062328 (2018)], and find it to be a promising alternative to the conventional grid architecture.
We also show that the lower bound that this metric places on the creation time of highly entangled states can be saturated with a constructive protocol, up to a factor logarithmic in the number of qubits.
\end{abstract}

\maketitle

\section{Introduction}
As the development of quantum computers progresses from the construction of qubits to the construction of intermediate-scale devices, quantum information scientists have increasingly begun to explore various architectures for scalable quantum computing \cite{Monroe2013,Ahsan2015, Pirker2018, Villalonga2018}. 
Researchers have quantified the cost imposed by moving from one architecture to another \cite{Cheung2007,Holmes2018} and optimized the placement of qubits on a fixed architecture \cite{Rosenbaum2010,Rosenbaum2012,Pedram2016}. Experimentalists have also begun to test different architectures in laboratory settings \cite{Linke2017, Maslov2017}.

In this work, we are interested in developing tools to evaluate the relative performance of different architectures. Here, ``architecture'' refers to the connectivity graph that defines the allowable elementary operations between qubits. We propose a natural metric based on entanglement measures. When several physical models are represented by a graph $G=(V,E)$, with a set of vertices $V$ corresponding to qubits, and a set of weighted edges $E$ corresponding to two-qubit operations (where the weights denote the maximum rates of operations), a useful metric is given by what we dub the ``rainbow time,'' 
\begin{equation}
	\tau_\rb (G) = \max_{F \subset V, \abs{F} \leq \half \abs{V}} \frac{ \abs{F}}{\abs{\partial F}}, \label{eqn:rainbowtime}
\end{equation}
where $|\partial F|$ denotes size of the boundary of $F$, i.e.\ the total weight of edges connecting $F$ and $\bar{F}=V- F$.

We show that the rainbow time is a lower bound on the time required to create a highly entangled state on the graph (i.e., states of $N$ qubits with $\mathcal{O}(N)$ bipartite entanglement). It is also the reciprocal of a well-studied graph quantity known as the isoperimetric number\,\cite{Mohar1989}.
We note that this lower bound holds even when measurement and feedback are allowed to speed-up entanglement generation, such as in the case of Greenberger-Horne-Zeilinger states\,\cite{Meignant2018}.
In contrast to Ref.~\cite{Bapat2018}, where architectures are evaluated assuming that only unitary operations are permitted, our results apply to the more general setting that allows non-unitary operations.

As a complementary result, we show that this lower bound is nearly tight -- a procedure that distributes Bell pairs using maximum-flow algorithms nearly saturates this bound to produce $\mathcal{O}(N)$ entanglement across any bipartition, up to $\mathcal{O}(\log N)$ overhead.
This suggests that beyond providing a bound, the rainbow time would be a useful witness to the speed at which entanglement can actually be generated.

\section{Physical Model}
In this paper, we evaluate the performance of quantum architectures with a connectivity graph given by $G$.
Each vertex in the graph represents a single data qubit, and an edge exists between two vertices if two-qubit operations can be performed between them. We interpret the edge weight  $w_{ij}$ between vertices $i$ and $j$ as representing bandwidth, so that higher-weighted edges are capable of performing more two-qubit operations in a single unit of time.

\begin{figure}[tb]
	\centering
	\includegraphics[width=0.8\linewidth]{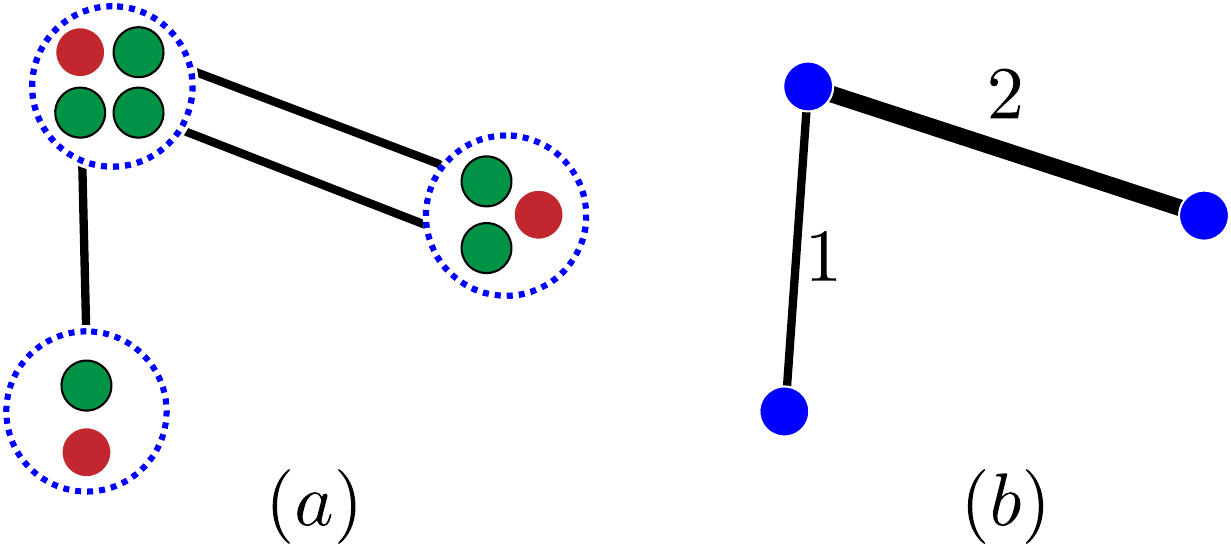}
	\caption{Illustration of how a model with ancilla mediator qubits can be abstracted into one in which only data qubits and edge weights are tracked. In panel (a), each module (blue dashed circle) contains one data qubit (red) and several ancilla mediator qubits (green) that form Bell pairs with other modules. In panel (b), the module as a whole is represented by blue circles, while the ancilla mediator qubits are now represented by edge weights. Only the states of the data qubits are tracked.}
	\label{fig:abstraction}
\end{figure}

We consider an example physical model where the edge weights represent the rate of distribution of entangled pairs as in Ref.~\cite{Chou2018}.
Each vertex is a small module that contains a data qubit and some ancilla qubits.
In each unit of time, Bell pairs are generated between the ancilla qubits on the edges of the graph, which can then be used to perform two-qubit gates on the data qubits \cite{Gottesman1999,Jiang2007}.
The process of moving from this model to an abstracted connectivity graph is illustrated in \cref{fig:abstraction}.  
We assume that measurements, classical communication, and intra-module unitaries are arbitrarily fast, such that the bottleneck is given by quantum operations between modules.
For example, this model can describe a trapped-ion system which uses photonic interconnects to generate entanglement between modules as in Refs.~\cite{Brown2016, Nigmatullin2016}.
In this framework, vertex degrees and total graph edge weights represent required ancilla overheads, justifying their use as cost functions in Ref.~\cite{Bapat2018}.

While, for simplicity, we will focus in the main text on the above model, our results also apply to other physical models, up to constant-factor overheads.
For example, since any two-qubit operation between data qubits can be performed by consuming two Bell pairs\,\cite{Eisert2000}, the above model is equivalent to a model where edge weights are proportional to rates of two-qubit operations.
In Appendix~\ref{appx:models}, we show in more detail how to extend our results to this model, as well as to a model where edge weights represent coupling strengths in a Hamiltonian.

\section{Entanglement capacity}
Given a graph $G$, we wish to bound the total possible increase in a given entanglement measure after $n$ rounds of entanglement distribution over its links.
Suppose we fix a bipartition of the graph into two subgraphs supported on vertex subsets $F$ and $\bar{F}$.
We consider a general entanglement measure, $S$, which quantifies the bipartite entanglement between $F$ and $\bar{F}$.
We assume the following axioms: $S$ is zero for product states $\rho_F \otimes \rho_{\bar{F}}$, additive between non-entangled regions, $S \left( \rho_{F\bar{F}} \otimes \tau_{F \bar{F}} \right)$\,\,=\,\,$S(\rho_{F\bar{F}}) + S(\tau_{F \bar{F}})$, and non-increasing under local operations and classical communication. Entanglement measures that obey these axioms include the entanglement cost, the distillable entanglement, and the entanglement of formation \cite{Bennett2003,Horodecki2009}. All of these measures are identical to the von Neumann entropy for pure states.

By the result of Ref.~\cite{Bennett2003}, the entanglement after $n$ rounds is bounded by $n$ times the maximum single-round entanglement. We will therefore bound the entanglement generated in one round, going from $\rho$ to $\rho'$. To produce $\rho'$, we begin with $\rho$ and then generate entanglement on the graph edges. This means that $w_{ij}$ ancilla Bell pairs are generated for each edge $(i,j)$ crossing the boundary $\partial F$. The total number of Bell pairs is therefore $\abs{\partial F}$, the sum over all the weights, 
\begin{equation}
	\abs{\partial F} = \sum_{i \in F, j \in \bar{F}} w_{ij}.
\end{equation}
Ignoring ancillas purely local to $F$ or $\bar{F}$, the resulting state is $\rho \otimes \rho_\mathrm{Bell}^{\otimes \abs{\partial F}}$. The final state $\rho'$ is then generated by local operations, assisted by classical communication, on this state. We denote the state that results from an arbitrary round of local operations and classical communications on $\rho$ as $\mathrm{LOCC}(\rho)$. Therefore, our axioms for $S$ allow us to write:
\begin{align}
	S\left( \rho' \right) &= S \left( \mathrm{LOCC} \left( \rho \otimes \rho_\mathrm{Bell}^{\otimes \abs{\partial F}} \right) \right) \nonumber \\
	&\leq S \left( \rho \otimes \rho_\mathrm{Bell}^{\otimes \abs{\partial F}} \right) \nonumber \\
	&= S(\rho) + \abs{\partial F} S\left( \rho_\mathrm{Bell} \right), \nonumber \\
	\implies S(\rho') - S(\rho) &\leq \abs{\partial F} S\left( \rho_\mathrm{Bell} \right).
\end{align}
Working in the units of $S(\rho_{\rm Bell})=1$, we refer to this upper bound on the change in entanglement, $\Delta S \leq \abs{\partial F}$, as the \textit{entanglement capacity} of the $(F, \bar{F})$ bipartition in the graph $G$.

\section{Rainbow states}
We now define a highly entangled state whose creation serves as a benchmark for the performance of a quantum computing architecture.

Entanglement makes a useful benchmark for any quantum computer because it can be shown that computations that do not produce entanglement can be efficiently simulated classically \cite{Vidal2003, Verstraete2004}
\footnote{Although universal quantum computation is possible in the limit of vanishing entanglement by implementing any quantum circuit $\C$ in a way that's controlled by a qubit in the state $\sqrt{1-\epsilon}\ket{0}+\sqrt\epsilon\ket{1}$~\cite{VandenNest2013}, such computation still requires the ability to implement the circuit $\C$. This means that any entanglement-based bound on the time-complexity of implementing $\C$ would still apply to the $\epsilon$-entangled version.}.
Further motivation for producing highly entangled states can be found in quantum simulation, where a quantum simulator of general applicability ought to be capable of representing and simulating highly entangled states \cite{Cirac2012}.

To select a particular entangled state for benchmarking, we consider ``rainbow states.'' In 1D contexts, for even $N$, a rainbow state is one in which qubits $i$ and $N-i$ are maximally entangled \cite{Ramirez2015,Alexander2018a}. The state itself is maximally entangled across a bipartition between the first $N/2$ qubits and the rest.

\begin{figure}[tb]
	 \centering
	 \includegraphics[width=0.8\linewidth]{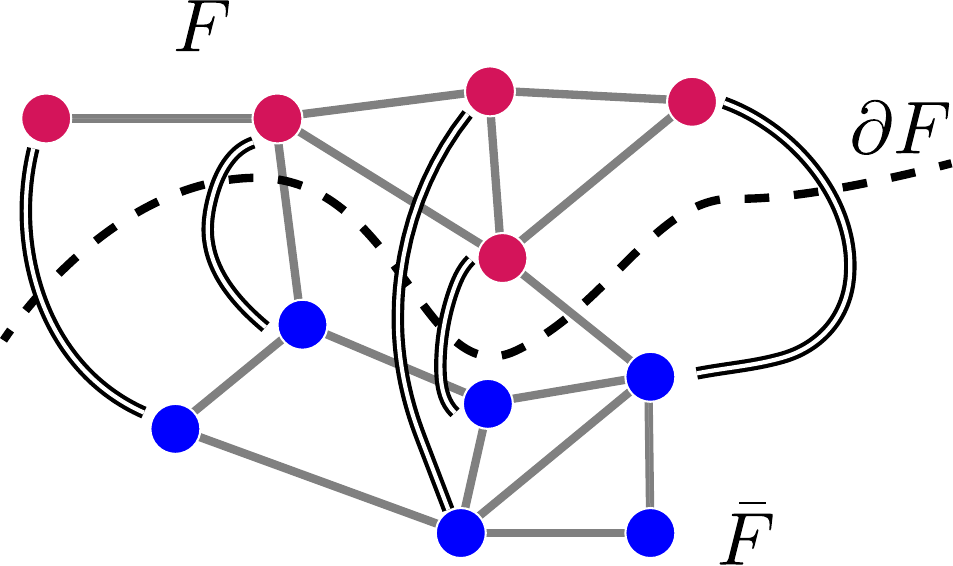}
	 \caption{An illustration of how a rainbow state is defined on an arbitrary subgraph $F$. Here, gray lines represent the connectivity graph of allowed two-qubit interactions, while doubled black lines represent maximally entangled qubit pairs. Qubits without a doubled line are assumed to be in state $\ket{0}$.}
	 \label{fig:bipartition}
	 \vspace{-9pt}
 \end{figure}

We extend this construction to arbitrary graphs. Suppose we consider a set of qubits $V$ and any subset $F$\,$\subset$\,$V$, with the requirement that $\abs{F}$\,$\leq $\,$\half \abs{V}$.
Denote by $F_i$ the $i$th vertex of $F$ using an arbitrary ordering, and similarly use $\bar{F}_i$ to index vertices in the complement $\bar{F}$.
We can then define a ``rainbow'' state as one in which qubit $F_i$ and qubit $\bar{F}_i$ form a Bell pair, and any additional qubits in $\bar{F}$ are left in the state $\ket{0}$.
This state is illustrated for a particular choice of $F$ and ordering in Fig.~\ref{fig:bipartition}.
Note that this construction is only well-defined if $\abs{F}$\,$\leq $\,$\half \abs{V}$, as otherwise there will not be enough data qubits in $\bar{F}$ to form Bell pairs with all the data qubits in $F$.
The arbitrary ordering allows multiple rainbow states to be defined from the same $F$.

\section{Rainbow times and isoperimetric number}
Using the model for quantum architectures in which each edge weight of a graph $G$ denotes the rate of entanglement generation across that edge, we can calculate the lower bound on the time required to create a rainbow state, according to the entanglement capacity. For any vertex subset $F$ we define this time as
\begin{equation}
	t(F) = \frac{ \abs{F}}{\abs{\partial F}} = \frac{\text{number of qubits in }F}{\text{entanglement capacity of } (F,\bar F)}.
\end{equation}
As we have shown, the entanglement capacity corresponds to the total weight of edges across the boundary, which constrains the amount of entanglement that can be distributed to the subsystem $F$ from its complement $\bar{F}$ in unit time.

Although there are many choices for a highly entangled physical state associated with the subset $F$ that would be hard to create, here we argue why the above metric $t(F)$ suffices for most considerations.
Although there are many different states with $\mathcal{O}(N)$ entanglement which could be used to evaluate graphs, the rainbow state is easy to conceptualize and create.
Since any bipartite entangled state can be converted either to or from Bell pairs through entanglement concentration or dilution\,\cite{Bennett1996a}, the rainbow state offers insight into the time required to create a general bipartite entangled state.
Furthermore, rainbow states arise as ground states of novel models in condensed-matter physics\,\cite{Zhang2017}, and thus the ability to create them can be important for quantum simulation.
The difficulty to create rainbow states is also recognized in Ref.~\cite{Meignant2018}.
While there is freedom in defining a physical rainbow state via the pairing of vertices in $F$  with those in $\bar{F}$, the precise choice of pairing does not affect the minimum time required to create the state according to the entanglement capacity, $t(F)$.
While different rainbow states that share a common subset $F$ may differ in how quickly they can be created, $t(F)$ serves as the common lower bound on the creation time for all of them, and thus we will focus on that metric here.

We will now use $t(F)$ to evaluate the quantum architecture $G$, the larger graph that contains $F$ as a vertex subset.
To do this, we find the maximum $t(F)$ given $G$.
Note that this is not the same as maximizing entanglement entropy, which would simply yield half the graph without any consideration of the graph structure.
Instead we ask: Of all the maximally-entangled states we can build by bipartitioning $V$ into $F$ and $\bar{F}$, which of them is slowest to build according to the entanglement capacity?
We call the associated quantity $t(F)$ the \textit{rainbow time} of the graph $G$ and denote it $\tau_\rb(G)$, as defined in \cref{eqn:rainbowtime}.

The rainbow time has a simple and attractive interpretation, can be directly connected to quantum computing tasks, and is applicable to various physical models of computation.
In addition, it can be directly connected to a quantity known as the isoperimetric number $h(G)$ \cite{Mohar1989}, sometimes also known as the Cheeger constant, which is well-studied in graph theory and computer science \cite{Mohar1988,Chung1998, Chung2005}. As we have defined it, the rainbow time is simply $\tau_\rb(G)$\,=\,$1/h(G)$ \footnote{Note that $\tau_\rb(G)$ can take on any nonnegative real value. In reality, the creation of a quantum state will always take an integer number of steps greater than or equal to one in our model. Therefore, $\ceil{\tau_\rb}$ can be used as a measure of the ``number of rounds required'' in cases where this is important.}. Thus, aiming to minimize the rainbow time (so that large entangled states can be easily created) in a quantum architecture is equivalent to maximizing the isoperimetric number. An ``isoperimetric set'' is a vertex subset $F$ that achieves $t(F) = \tau_\rb(G)$. Often, isoperimetric numbers appear in the context of expander graphs, which are constructed to possess large isoperimetric numbers \cite{Goldreich2011} and are used to prove important results in complexity theory \cite{Ajtai1983,Reingold2008,Dinur2007}. Intuitively, a small isoperimetric number (large $\tau_\rb$) means that a graph has bottlenecks, and a sizable subset can easily be disconnected by removing relatively few edges. This also implies that an architecture with large $\tau_\rb$ is more prone to becoming disconnected due to the failure of a small number of edges.

Even though computation of the exact rainbow time is NP-hard for general graphs\,\cite{Mohar1989}, it can be approximated to within an $\mathcal{O}(\sqrt{\log N})$ factor~\cite{ASV2009}.
There are also efficiently computable bounds on the rainbow time, including ones using the eigenvalues of the graph Laplacian\,\cite{Mohar1989}.
Furthermore, for many specific graphs, we can evaluate the rainbow time efficiently.
In Appendix~\ref{appx:graphs},
we have done this for the complete, star, and grid graphs, as well as the hierarchical products and hierarchies presented in Ref.~\cite{Bapat2018}. In particular, we compare hierarchies to $d$-dimensional grids and show that, for some parameters, hierarchies have lower rainbow time and lower total edge weight than grids, making them promising architectures for quantum computing.

\section{Creating rainbow states}
So far we have shown that rainbow time $\tau_\rb$ serves as a lower bound for generating maximum entanglement across any bipartition of the system.
We now examine whether this bound can be saturated, in the sense that one can create a rainbow state across any bipartition in time $\tilde{\mathcal{O}}(\tau_\rb)$.
We will show that for a general graph, there is an explicit protocol that prepares a rainbow state in time no more than $\lceil\tau_\rb \ln \abs{F} \rceil$ for any bipartition where $F$ is the smaller subset, indicating that the bound $\tau_\rb$ is tight up to a logarithmic factor.

\begin{figure}[tb]
	\centering
	\includegraphics[width=\linewidth]{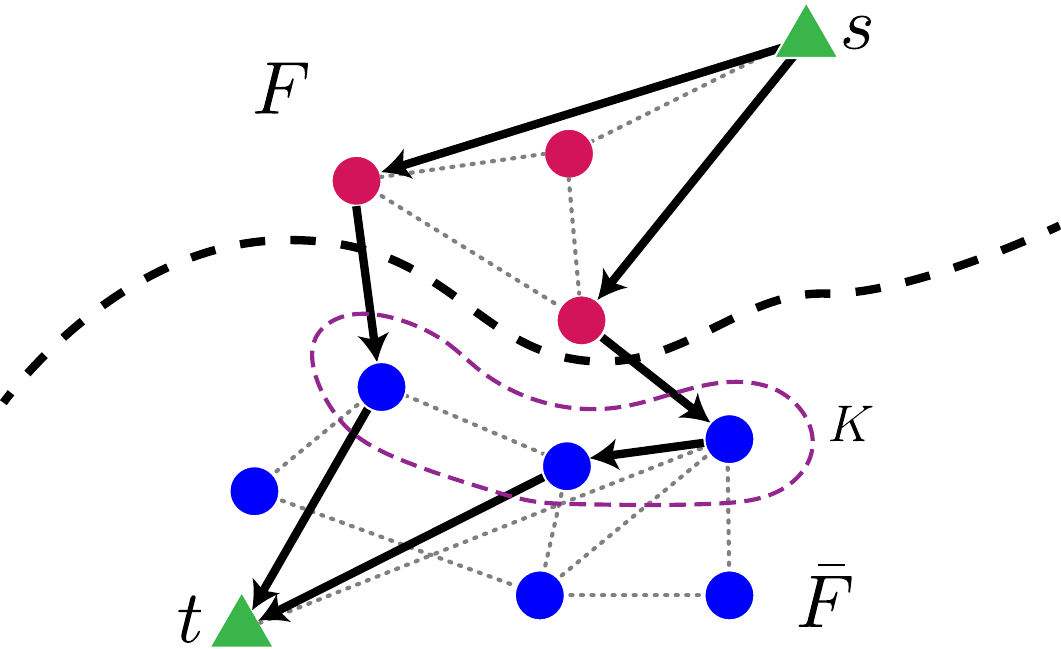}
	\caption{An illustration of the fictitious nodes added to the isoperimetric set, $F$ and a set of equal size $K$ (encircled by purple dashed line), to create a flow network. The new fictitious nodes, $s$ and $t$, appear as green triangles connected to every node in $F$ and $K$ respectively; the original nodes and edges are pink (in $F$) and blue (in $\bar{F}$) circles. The edges have weight one. The flow, shown by arrows, transfers $\ceil{\abs{F}/\tau_\rb} = 2$ units of entanglement across the bipartition. Gray, dotted edges are not used by the flow.
        }
	\label{fig:flow}
\end{figure}

We begin the proof by mapping the problem of creating rainbow state to the MaxFlow problem in computer science\,\cite{Elias1956}.
Here, we restrict our attention to quantum architectures on graph $G=(V,E)$, where the edge weights are integers that represent the number of Bell pairs that can be generated across the edge per unit time.
Suppose we are given arbitrary vertex subsets $F$ and $K$, where $|F|=|K|\le |V|/2$, and $K$\,$\subset$\,$\bar{F}$.
To create a Bell state between a given pair of nodes in a single time step, we can specify a path connecting them on the graph $G$,
generate Bell pairs on each edge along that path, and then perform entanglement connection on each internal node to convert the string of Bell pairs into one long-distance Bell pair.
We can create many distant Bell pairs in this way during a single time step by specifying many paths.
However, the set of paths must not use any edge more often than the weight of that edge allows for, since by definition the weight of an edge limits the number of Bell pairs the edge can generate in a unit time step.
Thus, we can interpret the weight of each edge as its \emph{capacity}, and the collection of paths as a \emph{flow} of entanglement from $F$ to $K$, as illustrated in Fig.~\ref{fig:flow}.
Suppose we now attach a fictitious source node $s$ to each node in $F$, and a fictitious sink node $t$ to every node in $K$.
Then the problem of maximizing the number of Bell pairs simultaneously generated between $F$ and $K$ is the same as the problem of maximizing the flow from the source $s$ to the sink $t$.
The latter problem is known as MaxFlow, visualized in Fig.~\ref{fig:flow}, and an explicit protocol to give the maximum possible amount of flow can be found efficiently via e.g., the Ford-Fulkerson algorithm \cite{Ford1956}. Note that if all the edge weights are integers, a flow of maximum value exists in which the flow carried by each edge is also an integer \cite{Fulkerson1958}.

To demonstrate that a flow approach yields an efficient creation of a rainbow state, we invoke the MaxFlow-MinCut theorem, which says that the maximum flow has the same value as the minimum cut\,\cite{Elias1956}. 
Here, a ``cut'' means a bipartition of the graph separating $s$ and $t$, and its value is the total weight of all edges that cross the bipartition.
By finding a lower bound on the value of all possible cuts in a graph, we show that a flow larger than or equal to this bound must exist.

Suppose that we now consider any cut of the graph into some arbitrary pair of subsets $\{s\} \cup S$ and $\{t\} \cup T$. The boundary of this cut will consist of edges from $s\to T$, $S\to t$, and $S \to T$.
Its magnitude can be written as
\begin{align}
	\abs{\mathrm{Cut}(S,T)} &= \abs{T \cap F} + \abs{S \cap K} + \abs{\partial S},
	\label{eqn:cutbound}
\end{align}
since $s$ and $t$ are connected only to nodes in $F$ and $K$, respectively, and the edges in $S \to T$ are just the boundary of $S$ in the original graph.
To evaluate $\abs{\partial S} = \abs{\partial T}$, we will assume that $\abs{S} \leq \half \abs{V}$, meaning we can apply the isoperimetric condition $\abs{S} \leq \abs{\partial S} \tau_\rb$. (If this is not the case, then a near-identical argument can be made applying this condition to $T$.) To account for cases where $\tau_\rb < 1$, we will write this as $\abs{\partial S} \geq m \abs{S}$ where $m = \min \left(1 , 1 /\tau_\rb\right)$. We then note that:
\begin{align}
	\abs{\partial S} &\geq m \abs{S} \geq m \left( \abs{S \cap F} + \abs{S \cap K} \right) \nonumber  \\ &\geq  m \left( \abs{F} - \abs{T \cap F} + \abs{S \cap K} \right). 
\end{align}
By inserting this lower bound for $\abs{\partial S}$ into Eq.~\eqref{eqn:cutbound}, we obtain
\begin{equation}
	\abs{\mathrm{Cut}(S,T)} \geq \left(1 - m \right) \abs{T \cap F} + \left(1 + m \right) \abs{S \cap K} + m \abs{F}.
\end{equation}
Since we know $m \leq 1$, we obtain the final bound on the cut magnitude,
\begin{equation}
	\abs{\mathrm{Cut}(S,T)} \geq m \abs{F}.
\end{equation}

If $m = 1$ (i.e., $\tau_\rb$\,$\le$\,$1$), then it follows that the value of the smallest cut is greater than $\abs{F}$, meaning that a flow exists of magnitude at least $\abs{F}$, which creates the rainbow state in a single round. If $m < 1$ (i.e., $\tau_\rb$\,>\,$1$), then we find that a flow exists of magnitude $\abs{F}/\tau_\rb$ \footnote{Since it is guaranteed to be integer-valued for graphs with integer-valued edge weights, the flow must in fact be of magnitude $\ceil{\abs{F}/\tau_\rb}$, but this makes no difference to the argument.}. Once $\abs{F}/\tau_\rb$ nodes are entangled, they can be disconnected from $s$ and $t$, and the process repeated on a new set of nodes $F_1 \subset F$. Therefore after $n$ rounds of computation, the remaining set of nodes waiting for entanglement $F_n$  is produced by removing $1/\tau_\rb$ of the nodes in set $F_{n-1}$, with $F_0$\,=\,$F$, allowing us to compute the maximum size of $F_n$ inductively:
\begin{align}
	\abs{F_n} &\leq \left( 1 - \frac{1}{\tau_\rb} \right) \abs{F_{n-1}} \nonumber\\
	&\leq \left( 1 - \frac{1}{\tau_\rb} \right)^n \abs{F} < e^{- n/ \tau_\rb} \abs{F}.
\end{align}
Once $\abs{F_n} < 1$, the process is complete, as there are no fractional nodes. It follows that $\lceil \tau_\rb  \ln \abs{F} \rceil$ rounds suffice to complete the entangling process.

\section{Outlook}
In this work, we have presented a new metric for evaluating proposed architectures for quantum computers. 
While we have proven that any vertex subset $F$ can have a rainbow state prepared in $\lceil \tau_\rb   \ln \abs{F} \rceil$ time, test simulations on many example small graphs suggest that flow-based algorithms can create rainbow states in $\lceil\tau_\rb\rceil$ time. It is thus possible that the logarithmic factor can be removed and that the rainbow time lower bound is fully tight and saturable. 
In addition, although our argument suggests that for any bipartition of the system, \textit{there exists a} rainbow state that can be created in $\lceil\tau_\rb \ln \abs{F}\rceil$ time, other rainbow states (where the connections between node pairs are permuted) may take longer.
It would be interesting to upper bound the creation time of arbitrary rainbow states using tools from classical network theory such as routing time\,\cite{Schoute2016, Childs2019}. 

Finally, another open question is how the entanglement capacity, used here in terms of the rainbow time, can be applied to the analysis of quantum algorithms. 
While the rainbow time is not enough to provide an upper bound on the time-complexity of running a quantum algorithm on a given quantum architecture, it can provide a lower bound when the amount of entanglement required in the algorithm is known.
References~\cite{Orus2004,Kendon2006} explore the question of how entanglement grows during Shor's algorithm and in adiabatic quantum computing.
These complement other results showing that low-entanglement systems can be simulated efficiently on a classical computer\,\cite{Vidal2003,Schuch2008}. Rainbow time can also be used to benchmark algorithms for compilation and gate decomposition of quantum circuits, by comparing their realized circuit depth to this theoretical minimum required time.

\begin{acknowledgments}
We thank A.~Childs, A.~Harrow, L.~Jiang, D.~Leung, G.~Smith, and X.~Wu for discussions.
Z.E., A.B., J.R.G., A.D., and A.V.G.\ acknowledge funding by ARO MURI, DoE ASCR Quantum Testbed Pathfinder program (Award No.\ DE-SC0019040), AFOSR, ARL CDQI, NSF PFCQC program, AFOSR MURI, DoE BES Materials and Chemical Sciences Research for Quantum Information Science program (Award No.\ DE-SC0019449), DoE ASCR Accelerated Research in Quantum Computing program (Award No.\ DE-SC0020312), and NSF PFC at JQI\@. A.B.\ is supported in part by the QuICS Lanczos Fellowship. Z.E.\ is supported in part by the ARCS Foundation.  L.Z.\ is supported in part by the National Science Foundation and the Center for Ultracold Atoms. J.R.G.\ was supported in part by the NIST NRC Research Postdoctoral Associateship Award. F.C.\ is funded in part by EPiQC, an NSF Expedition in Computing, under grant CCF-1730449; in part by STAQ, under grant NSF PHY-1818914; and in part by DOE Grants DE-SC0020289 and DE-SC0020331.
This work was performed in part at the Aspen Center for Physics, which is supported by National Science Foundation Grant No. PHY-1607611.
\end{acknowledgments}

\appendix

\section{Entanglement Capacities on Various Physical Models~\label{appx:models}}
In this Appendix, we will derive the entanglement capacity for several different physical models that can correspond to a graph. Consider a graph, $G$, and select a subset of the vertices, $F$. We then want to show that the maximum amount of entanglement that can be created between $F$ and $\bar{F}$ in unit time is proportional to the size of the boundary, $\abs{\partial F}$. We will allow arbitrary constant factors, and discuss how this bound arises in two different physical situations. As in the main text, we consider entanglement measures $S$ on two regions so long as $S$ obeys the following rules:
\begin{enumerate}
    \item Additively distributive over the tensor product, so $S(\rho \otimes \sigma) = S(\rho) + S(\sigma)$ if $\rho$ and $\sigma$ are supported on both sides of the bipartition.
    \item Zero for states which are a product of states on each region, $S ( \rho_F \otimes \rho_{\bar{F}} ) = 0$.
    \item Non-increasing after any operation which is local to each region, even if we permit classical communication.
\end{enumerate}
In the main text, we showed how to apply these axioms to the analysis of a case in which computation was performed by the production and consumption of Bell pairs. Here we also look at a gate model of computation and a case in which the graph describes the limits on a time-dependent interaction Hamiltonian.
\\

\textbf{Unitaries. }
In this model, each graph edge of weight $w_{ij}$ represents the capability to perform $w_{ij}$ unitaries between qubits $i$ and $j$ in a time step. These unitaries are freely chosen by the experimenter. For two qubits, the ability to apply multiple unitaries is no different from the ability to apply an arbitrary unitary. However, we are considering cases where the qubits are part of a larger system, meaning we may wish to perform unitaries in sequence on different pairs to perform a more complicated computation.

We note that every two-qubit unitary can be performed using two Bell pairs as a shared resource and applying local operations. This can be easily seen in the following process:
\begin{enumerate}
    \item Alice and Bob start with a data qubit each and two Bell pairs shared between them. They wish to implement an arbitrary two-qubit unitary using only local operations and classical control.
    \item Alice uses one Bell pair and classical communication to teleport her qubit to Bob.
    \item Bob uses his local operations to perform the desired two-qubit gate.
    \item Bob teleports Alice's qubit back to her.
\end{enumerate}
Therefore, the state $\rho'$ can be obtained from the state $\rho$ by using local operations and classical communication (LOCC) and consuming up to $2 \abs{\partial F}$ Bell pairs in the process. Since LOCC cannot increase $S$, it follows that:
\begin{align}
    S(\rho') &\leq S\left( \rho \otimes \rho_\mathrm{Bell}^{\otimes 2 \abs{\partial F}} \right) \\
    \implies \Delta S &\leq 2 \abs{\partial F} S(\rho_\mathrm{Bell}).
\end{align}
This suggests that the ability to perform arbitrary unitaries is up to twice as powerful as the ability to distribute arbitrary Bell pairs, which makes sense, as an arbitrary two-qubit gate cannot necessarily be performed with one Bell pair (for instance, SWAP requires two) \cite{Eisert2000}. Two Bell pairs however suffice to implement any arbitrary two-qubit unitary. In any case, this still yields an entanglement capacity $\Delta S = \mathcal{O}(\abs{\partial F})$ bound as desired.
\\

\textbf{Hamiltonians. }
We will now consider a case in which the graph describes a Hamiltonian, possibly time-dependent. The graph will restrict the strength of these Hamiltonians. If we assume that $G = (V, E)$, then the Hamiltonian can be written as a sum over the two-qubit operations:
\begin{equation}
\label{eq:Hgeneralform}
    H(t) = \sum_{(i,j) \in E} h_{ij} (t).
\end{equation}
We then impose the condition:
\begin{equation}
    \forall t: \| h_{ij}(t) \| \leq w_{ij},
\end{equation}
where $w_{ij}$ is the $i$-$j$ edge weight. 
We can then apply the ``small incremental entangling" (SIE) theorem \cite{VanAcoleyen2013}. In particular, we apply the special case used in Ref.~\cite{Gong2017} to bound the total amount of entanglement generated by this Hamiltonian. If $H$ is a sum of pairwise Hamiltonians $h_{ij}$ acting on qubits, then the time-rate of entanglement generation on a set $F$ of sites is:
\begin{equation}
    \abs{\frac{ \mathrm{d} S_F }{\mathrm{d} t} } \leq 36 \log(2) \sum_{i \in F, j \in \bar{F}} \| h_{ij} \|.
\end{equation}
Here, $S_F$ is the von Neumann entropy of the reduced density matrix on the region $F$. This can be derived from Eq.~(3) of Ref.~\cite{Gong2017}, and specifying two-body terms and qubit sites, but the result could be extended to qudits or general $k$-body interactions. The sum over Hamiltonian norms, in the graph context, corresponds to a sum over graph edges. Since every Hamiltonian strength is limited by the corresponding edge weight, $\sum \| h_{ij} \| \leq \sum w_{ij} = \abs{\partial F}$. Therefore, we can specifically say that for this case, $\Delta S_F = \mathcal{O}( \abs{\partial F} )$. Many other entanglement measures, such as entanglement of formation or entanglement cost, can be related to the von Neumann entropy \cite{Horodecki2009}. In particular, many entanglement measures on mixed states can be defined as a weighted sum over pure state components; since none of the pure states can increase dramatically in entanglement under this process, the entanglement measure on the mixed state is similarly limited.


\section{Application to Hierarchical Product and Hierarchies\label{appx:graphs}}
\begin{figure}[tb]
	\centering
	\includegraphics[width=.45\textwidth]{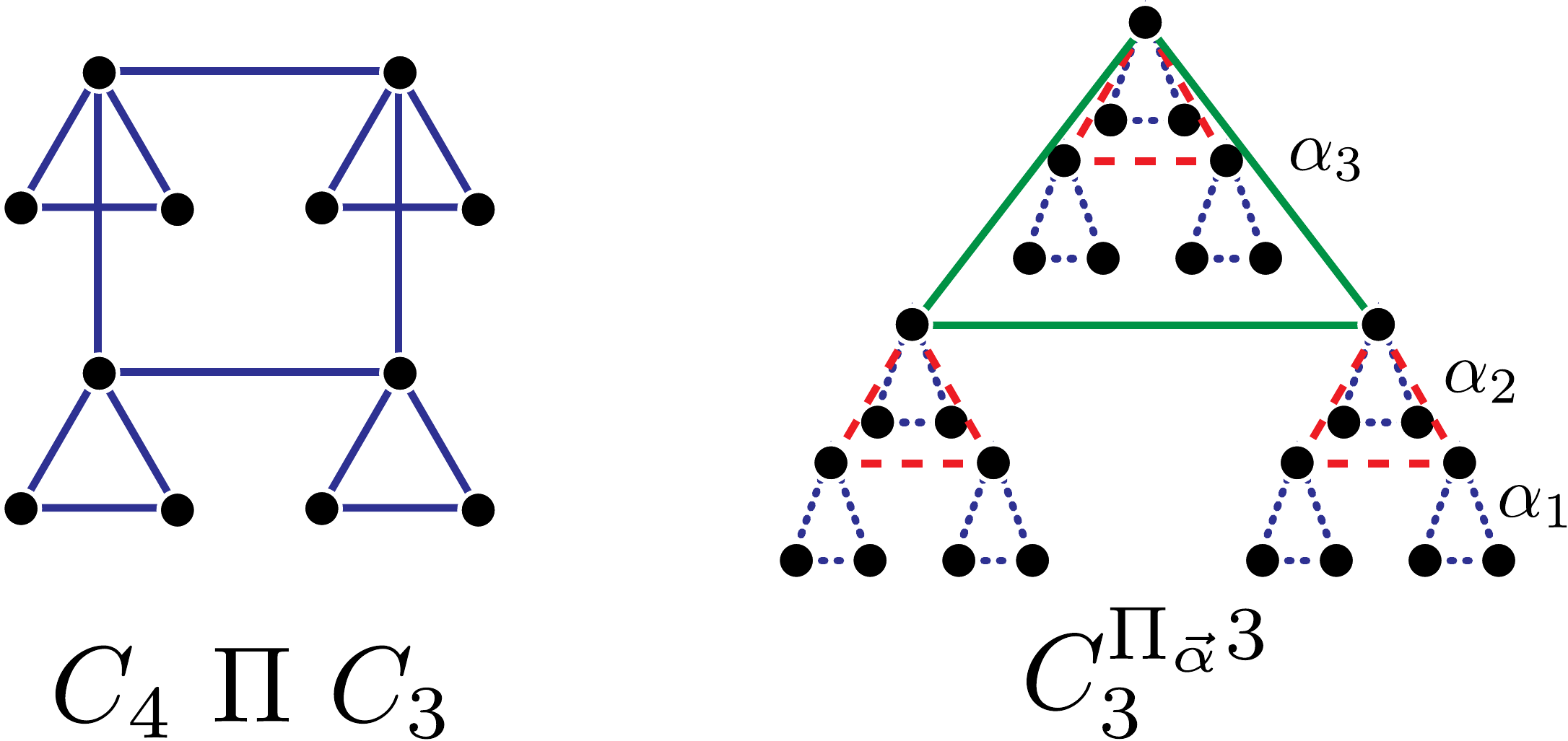}
	\caption{Examples of a hierarchical product (left) and a weighted hierarchy (right).}
	\label{fig:hierdemo}
\end{figure}

In this Appendix, we calculate the rainbow times for the hierarchical products and hierarchies of Ref.~\cite{Bapat2018}. A hierarchical product is a graph product denoted $G \uhp H$ in which $\abs{G}$ copies of $H$ are connected at their root (first) vertices by the graph $G$. By iterating this process, we can create a hierarchy, in which higher-level graphs connect lower-level identical sub-hierarchies. We also extend this concept to that of a weighted hierarchy, in which the edges on level $i$ have weight $\alpha_i$. We write a $k$-level hierarchy with a vector of weights $\vec{\alpha}$ as $G^{\vhp k}$, where $G$ is the base graph. Finally, if $\alpha_i = \alpha^{i-1}$, so that edge weight scales geometrically with the level of the hierarchy, then we simply write $G^{\whp k}$. Some examples are shown in \cref{fig:hierdemo}.

\begin{figure}[tb]
	\centering
	\includegraphics[width=.45\textwidth]{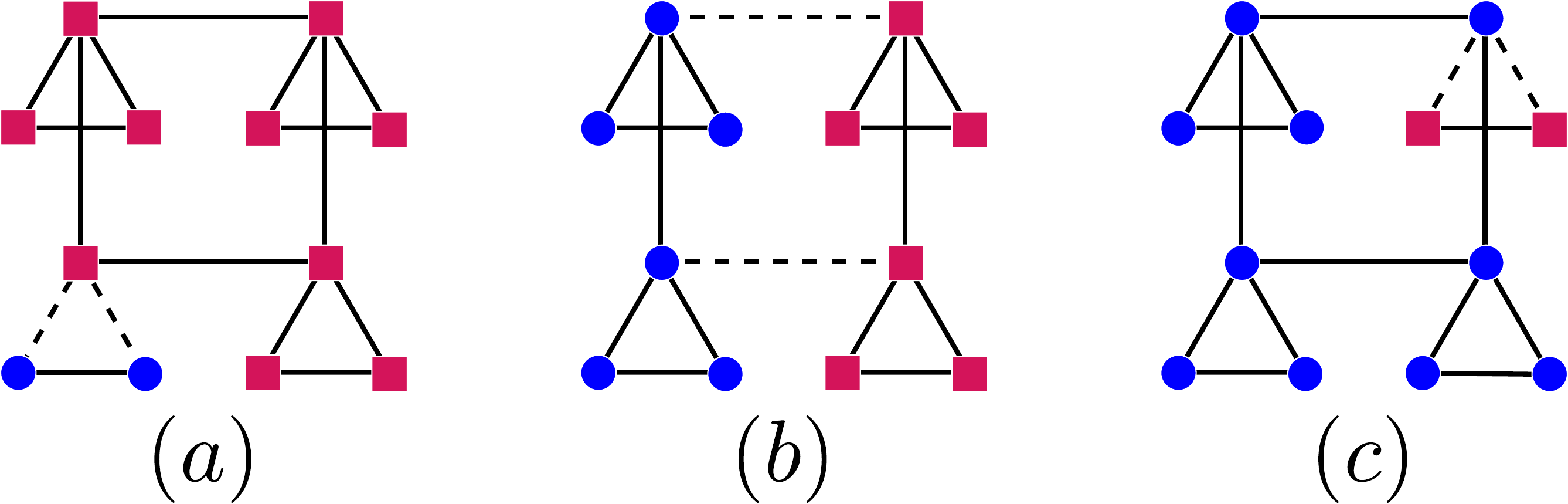}
	\caption{Three classes of subgraph used in our proof. Circles represent vertices in $F$, squares are vertices in $\bar{F}$, and dashed lines are edges in $\partial F$. (a) A situation in which part of one copy of $H_2$ is in $F$. (b) A situation in which the division between $F$ and $\bar{F}$ lies entirely in $H_1$. (c) A situation in which all but one of the copies of $H_2$ are entirely contained in $F$.}
	\label{fig:hp_strategies}
\end{figure}

To calculate the rainbow time for a hierarchical product, we make use of the result from Ref.~\cite{Mohar1989} that there must exist an isoperimetric set [a vertex set $F$ such that $\tau (F) = \tau_\rb(F)$] that is connected and whose complement $\bar{F}$ is connected. Therefore, we will look at all possible subgraphs of $H_1 \uhp H_2$ where both $F$ and $\bar{F}$ are connected. From these, we will search for the one with the largest $\tau (F)$. Since some isoperimetric set is guaranteed to exist in this set of subgraphs, this maximization over $\tau (F)$ in this set will also give us $\tau_\rb(H_1 \uhp H_2)$. We will begin by specifying three cases, illustrated in \cref{fig:hp_strategies}. These cases cover all possible subsets with the right connectedness properties and therefore allow us to find the maximizing set for the graph and $\tau_\rb (H_1 \uhp H_2)$. 

One such set would cover part of one copy of $H_2$. However, note that if the root vertex of $H_2$ were included in $F$, then we would have to include all the descendants of $H_2$, since otherwise $\bar{F}$ would not be connected. Therefore, this class will only include subsets of $H_2$ which do not include the root vertex. In this case, we must maximize over all possible subsets of $H_2$ to find the maximum $\tau (F)$. This may seem like it would yield $\tau_\rb(H_2)$; however, in this instance we can pick subsets of $H_2$ which make up a majority of $H_2$, which is not allowed for $\tau_\rb$. We define the \textit{unrestricted rainbow time} as 
\begin{equation}
	u_\rb (G) = \sup_{F \subset G \setminus G_1} \tau (F).
\end{equation}
Here, $G \setminus G_1$ refers to $G$ with its first vertex removed. Therefore, any set from this class will offer a candidate rainbow time of at most $\tau (F) = u_\rb(H_2)$.

The second class of candidate sets would cross one or more copies of $H_2$. Since $F$ must be connected, the path between these copies must be included in $F$, which means the root vertices of each $H_2$ that connect to each other via $H_1$ must also be in $F$. Then, as shown above, the entire copy of $H_2$ must be included. As a result, this case is equivalent to choosing copies of $H_2$ and either entirely including them in $F$ or entirely excluding them. This problem reduces to dividing up $H_1$, and then calculating as if each vertex had an effective volume of $\abs{H_1}$. Therefore, we can find the maximum $\tau (F)$ of these sets by simply finding $\tau_\rb (H_1)$ and scaling it by $\abs{H_2}$.

The final class of sets $F$ which meets the connectedness criteria would be an $F$ which includes all of $H_1$ and then all but one copy of $H_2$ completely, with perhaps some of the remaining $H_2$ also included. However, this $F$ would necessarily be larger than half of the total graph $H_1 \uhp H_2$, and therefore we can discard it as a candidate set for determining the rainbow time. We combine the first two options and conclude that:
\begin{equation}
	\tau_\rb ( H_1 \uhp H_2) = \max \left( u_\rb (H_2) , \abs{H_2} \tau_\rb (H_1) \right).
\end{equation}

We now seek to apply this to hierarchies $G^{\vhp k}$. Just as before, if a vertex is included in $F$, then we must also include in $F$ all its descendants in the hierarchy, otherwise the complement $\bar{F}$ will not be connected. Therefore, all bipartitions can be reduced to choosing a particular level of the hierarchy to cut -- on that level, either a vertex will be included or not included, and this must apply to all of its descendants as well. Every bipartition can then be mapped to a bipartition of $G$, but one where every vertex is scaled by $\abs{G}^{i-1}$ due to the size of each sub-hierarchy [note that the large number of vertices not in $F$ do not contribute to $\tau (F)$]. In addition, $\tau (F)$ must also be modified by the edge weight, which we define to be $\alpha_i$ on level $i$.

There is one important difference between the top ($k$th) level and all others, which arises from the constraint that $\abs{F} \leq \half \abs{G^{\vhp k}}$. A cut on the top level must not include more than half of the highest-level copy of $G$, while all lower levels can use any cut at all as long as it does not include the root vertex. Whatever level we cut, the cut depends only on the base graph $G$, with each node standing for $\abs{G}^{i-1}$ total nodes below it. Therefore, we can write the overall $\tau_\rb$ as a maximization over these options:
\begin{equation}
	\tau_\rb (G^{\vhp k}) = \max \left( \frac{\abs{G}^{k-1}}{\alpha_k} \tau_\rb (G),\ \sup_{i < k}  \frac{\abs{G}^{i-1}}{\alpha_i} u_\rb (G) \right).
	\label{eqn:hiermax}
\end{equation}

For specificity, we will evaluate the case where $G = K_n$, the complete graph, and $\alpha_i = \alpha^{i-1}$, which was proposed in Ref.~\cite{Bapat2018} as an architecture. Here, the maximization over lower levels [the second term in \cref{eqn:hiermax}] can be reduced to either to the first level or the $k-1$ level, since we simply have to pick the largest element in a geometric sequence defined by $n/\alpha$. We can write the resulting maximization as a choice between three options,
\begin{equation}
	\tau_\rb ( K_n^{\whp k} ) = \max \left(1,\ \left( \frac{n}{\alpha} \right)^{k-1} \frac{2}{n},\ \left(\frac{n}{\alpha} \right)^{k-2}\right).
\end{equation}
Whereas one might have expected two options to arise (cut at the top or at the bottom), we actually have three. For $\alpha > n$, the edges grow in capacity too quickly for the increased volume to make a higher-level cut worthwhile, so the optimal cut is at the bottom, yielding a constant scaling with $n$. Two other options appear at $n > \alpha$, where cutting higher up the hierarchy allows for greater volume of qubits in $F$ without too much penalty caused by changing edge weights. The reason there are two strategies is that it may be possible to cut a larger portion of a lower hierarchy and exploit the split between $\tau_\rb$ and $u_\rb$. [For $K_n$ in particular, the cut that includes all but the root vertex satisfies $u_\rb (K_n)$.]

To place these results in context, we compare the rainbow time of $K_n^{\whp k}$ to the total rainbow time of other graphs. To do this, we write the rainbow time in terms of the total number of qubits in a graph, $N$, and concern ourself with the overall scaling.
For the purpose of comparison, we consider hierarchies where the number of levels scale logarithmically as $k=\log_n N$, while $\alpha,n$ are constant parameters independent of $N$.
In this language, $\tau_\rb(K_n^{\whp k}) = \Theta \left(N^{\max\left( 0, 1 - \log_n \alpha \right)}\right)$. We compare this to the rainbow time of some other graphs in \cref{tbl:graphstats}. References~\cite{Mohar1989,Chung1998} give the isoperimetric number for $K_N$, $S_N$ (the star graph of $N$ nodes), and grids (which are Cartesian products of paths). Satisfying sets for these graphs are: for $K_N$ and $S_N$, an arbitrary half of the nodes; for grids, a hypercube placed in one corner that takes up half the total volume.

One goal would be to identify a set of parameters where a hierarchy outperforms a $d$-dimensional grid architecture. We are most concerned with comparing to the $d$-dimensional grid because the other candidates we present, $K_N$ and $S_N$, both have very large degree, making them impractical for scalable architectures, although both have been used for small quantum devices \cite{Linke2017}. We find that the rainbow time of the hierarchy with base graph $K_n$ and scaling constant $\alpha$ will be better (smaller) than that of the grid if $\alpha > n^{(d-1)/d}$.
If it also holds that $n>\alpha$, then the hierarchy will accomplish this with a total edge weight scaling identically as the grid. It is possible to achieve a smaller pre-factor in this scaling under a suitable choice of $n, \alpha$; for example, when $d=2$, the choice of $n=3,4$ and $\alpha=n^{1-1/d}$ gives lower total edge weight for the hierarchy than the grid.
We conclude that a hierarchy $K_n^{\whp k}$ with $\alpha \in \left[n^{1 - 1/d}, n\right)$ has both lower rainbow time and lower total edge weight than a $d$-dimensional grid of qubits.

\begin{table}[tb]
	\begin{tabular}{l | r r r}
		Graph Name & $\tau_\rb$ & $w$ & $\Delta$ \\ \hline
		$K_N$ &$ N^{-1}$ & $N^2$ & $N$ \\
		$S_N$ & 1 & $N$ & $N$ \\
		$d$-dimensional Grid & $N^{1/d}$ & $N$ & $2d$ \\
		$K_n^{\whp k}$ & $N^{\max(0, 1-\log_n \alpha)}$ & $N^{\max(1, \log_n \alpha)}$ & $\log_n N$
	\end{tabular}
	\caption{Important statistics for graphs. Here, only the asymptotic scaling with $N$ is written. In addition to the rainbow time $\tau_\rb$ for each graph, we also include the total weight of all edges $w$, and the maximum graph degree $\Delta$. Rainbow times for graphs other than hierarchies can be found in terms of isoperimetric number in Refs.~\cite{Mohar1989,Chung1998}.}
	\label{tbl:graphstats}
\end{table}

\bibliography{References}

\end{document}